\documentclass[hyper,a4paper]{JHEP}
\usepackage{amsmath,graphicx,epsfig,amssymb,multirow}

\allowdisplaybreaks
\def\gsim{\raise0.3ex\hbox{$\;>$\kern-0.75em\raise-1.1ex\hbox{$\sim\;$}}}
\def\lsim{\raise0.3ex\hbox{$\;<$\kern-0.75em\raise-1.1ex\hbox{$\sim\;$}}}
\newcommand{\dimfour}{dim-4 }
\newcommand{\dimfive}{dim-5 }
\newcommand{\dimsix}{dim-6 }

\newcommand{\sff}{s_{ff}}

\preprint{ \\ IFIC/11-43 \\ \today}
\title{Effects of Anomalous Couplings on Spin Determination in Cascade Decays}
\author{Lisa Edelh\"auser$^{1,a}$, Werner Porod$^{1,2,b}$, Ritesh K. Singh$^{3,c}$ \\
$^1$Institut f\"ur Theoretische Physik und Astrophysik, Universit\"at W\"urzburg,\\
D-97074  W\"urzburg, GERMANY \\[1mm]
$^2$AHEP Group, Instituto de F\'{\i}sica Corpuscular \\
C.S.I.C./Universitat de Val{\`e}ncia \\
E--46071 Val{\`e}ncia, SPAIN \\[1mm]
$^3$Department of Physical Sciences\\
Indian Institute of Science Education \& Research Kolkata\\
Mohanpur 741252, West Bengal, INDIA
Email:~$^a$\email{ledelhaeuser@physik.uni-wuerzburg.de},\\
$^b$\email{porod@physik.uni-wuerzburg.de},
$^c$\email{ritesh.singh@iiserkol.ac.in} }

\abstract{Determining the spin of new particles is an important tool
for discriminating models beyond the Standard Model. We show
that in case of cascades of subsequent two body decays the existing strategy to
extract the spin from lepton and quark spectra can be used
without changes even if one allows for \dimfive  and \dimsix operators
which  might be induced by physics just beyond the reach of LHC.
We show analytically 
that these operators do not change the overall structure of
these spectra.  }

\begin{document}
%%%%%%%%%%%%%%%%%%%%%%%%%%%%%%%%%%%%%%%%%%%%%%%%%%%%%%%%%%%%%%%%%%%%%%%%%%%%%%%
%%%%%%%%%%%		INTRODUCTION				%%%%%%%%%%%%%%%
%%%%%%%%%%%%%%%%%%%%%%%%%%%%%%%%%%%%%%%%%%%%%%%%%%%%%%%%%%%%%%%%%%%%%%%%%%%%%%%

\section{Introduction}
The Large Hadron Collider (LHC) has started the direct exploration of
physics at the TeV scale and is searching for new particles which are
predicted in various extensions of the Standard Model (SM).  Many of these
models predict partners of the known SM particles, which usually have
the same quantum numbers and properties but for the mass and the
spin. For example, in supersymmetric models the fermions have scalar
partners whereas in models with extra dimensions fermionic partners
are predicted. Therefore, immediately after the discovery of new
particles the question will arise how to discriminate between the
various models and spin determination will play a crucial role here.
 
The question on how to determine the spin of new particles
has been addressed by several authors: in
\cite{Choi:2002jk,Alves:2008up,Osland:2008sy,Osland:2009tn,Eboli:2011bq}
$s$-channel resonances have been investigated and in
\cite{Barr:2005dz,Smillie:2005ar,Athanasiou:2006ef,Smillie:2006cd,%
  Wang:2006hk,Meade:2006dw,Kilic:2007zk,Alves:2007xt,Rajaraman:2007ae,Cho:2008tj,%
  Wang:2008sw,Burns:2008cp,Barr:2010zj,Chiang:2011kq} spectra of cascade decays of
subsequent two-body decays have been used to obtain information on the
spin.  An additional possibility to get information on the spin is
cross section measurements provided one knows the representation of
the particle produced \cite{Kane:2008kw}, e.g.~whether it is a colour
triplet or a colour octet. Also three-body decays have been
investigated in this direction for specific scenarios
\cite{Csaki:2007xm}.  In ref.~\cite{Edelhauser:2010gb} a strategy has
been worked out for scenarios where three-body decays are dominating.

For decay chains of at least three subsequent two-body decays
involving with at least  three visible SM particles in the final
state one can construct sufficient many kinematical variables
to determine the masses of four unknown particles even if
the lightest of them escapes detection, see e.g.~\cite{Barr:2010zj} and
references therein. This is in particular important if one requires
that a SM extension explains the observed dark matter (DM) due to
heavy weakly interacting particles, e.g.~the neutralino in supersymmetric
models or the lightest KK-excitation of the photon in models with
extra dimensions.
 However, these kinematical variables do not give any information
on the spin of the particles  involved whereas the angular distribution of
the decay products is highly effected by the spin of the intermediate
particle and each spin has its specific decay signature. This property
has widely been studied \cite{Barr:2004ze,Athanasiou:2006ef,Smillie:2006cd,Wang:2006hk}. Potential complications in
this context are at the one hand that one does not know a priori which
particles belong to which decay chain. On the other hand
it is usually also not known in which particular decay chain a
certain particle has been produced which can
potentially wash out effects as one has to sum over
different possibilities. An example of the later case
is in case of supersymmetric models the chain
\begin{equation}
\tilde q \to q \tilde \chi^0_2 \to q l^\pm \tilde l^\mp 
\to q l^+ l^-  \tilde \chi^0_1
\end{equation}
where $\tilde q$ is a scalar quark,  $\tilde l$ is a scalar lepton
and $\tilde  \chi^0_{1,2}$ are neutralinos. Here one can build 
kinematical variables using the jet originating from outgoing $q$ and the two
leptons \cite{Allanach:2000kt}. 

The fact, that LHC has failed so far to detect 
physics beyond the SM might imply that part of the corresponding
spectrum is somewhat beyond the reach of LHC, e.g.\ gluinos
or KK-excitations in the multi-TeV range. They might however
induce  \dimfive or \dimsix operators which are only mildly
suppressed which potentially endangers the above strategies
to determine the spin of new particles in cascade decays.
However, as we will demonstrate below, this is not the case
and one can use the strategies developed so far with changes
even if these additional operators are presented.
In the next section we will fix the notation and discuss the
potentially dangerous operators. In Section \ref{sec:resultsanomalous} we
derive the main results and then conclude in Section \ref{sec:discussion}.

%%%%%%%%%%%%%%%%%%%%%%%%%%%%%%%%%%%%%%%%%%%%%%%%%%%%%%%%%%%%%%%%%%%%%%%%%%%%%
%%%%%%%%%		BASIC IDEA & SETTINGS			%%%%%%%%%%%%%%%
%%%%%%%%%%%%%%%%%%%%%%%%%%%%%%%%%%%%%%%%%%%%%%%%%%%%%%%%%%%%%%%%%%%%%%%%%%%%%

\section{Setup and Kinematics}\label{sec:anomaloussub}

These methods rely on the fact that one can decompose the matrix element  of 
a scattering process $a+b\rightarrow i\rightarrow c+d$ such that the dependence of the scattering angle $\theta_{bc}$ is solely expressed in Wigner-d functions and a remaining reduced matrix element which only depends on the total angular momentum $J$, the helicities of the involved particles, and the particle masses. Using crossing-symmetry, one can rewrite the decomposition of this matrix element  \cite{Haber:1994pe,Leader:2001gr} for the s-channel process $a+b\rightarrow i\rightarrow c+d$  into the matrix element for the decay $a\rightarrow b+c+d$  which reads in the  rest frame of the intermediate particle $i$ \cite{Boudjema:2009fz}
\begin{align}
\mathcal{M}_{\lambda_c\lambda_d;\lambda_a\lambda_b}(s,\theta_{bc},\phi)&= (2s+1)~ d^s_{\lambda_i,\lambda_f}(\theta_{bc}) e^{i(\lambda_i-\lambda_f)\phi}\mathcal{M}^s_{\lambda_i\lambda_f}
\label{eq:wignerd}
\end{align}
with the differences of the initial and final particles' helicities $\lambda_i=(\lambda_a-\lambda_b),\lambda_f=(\lambda_c-\lambda_d)$  and where $\theta_{bc}$ is the scattering angle of two of the products. The Wigner-d rotation matrices $d^s_{\lambda_i,\lambda_f}$ depend polynomially on  $\cos\theta_{bc}/2$ with degree $2s$ \cite{brink}. After squaring, this leads to a polynomial in $\cos\theta_{bc}$ with degree $2s$
which gives us the spin-dependence of the angular distribution for this decay. Since we want to study frame-independent variables, we rewrite eq.~(\ref{eq:wignerd}) in terms of invariant mass $s_{ff}$ 
 of two visible decay products and the $\cos\theta_{bc}$ dependence of the matrix element translates into
\begin{align}d~\Gamma (X\rightarrow f\overline{f}Y) =d ~\sff\sum_{i}^{2s} b_i ~ \sff^{i}
\label{eq:maximalorder}
\end{align}
for the differential decay rate with maximal degree $2s$
where we have used the fact 
\begin{equation}
\sff\propto E_1E_2-||\vec{p}||\vec{q}|\cos\theta_{BC} 
\end{equation}
with $p,q,E_1,E_2$ being momenta and energies of the decay products.

We use use following notation for the masses and momenta in the investigated decays
\begin{align}
X(p,m_x)\rightarrow f(q_1,m_f) + I(p_I,m_I)\rightarrow f(q_1,m_f) +\overline{f}(q_2,m_f) +Y(q_3,m_y).
\label{eq:anomalousdecay}
\end{align}
where $X,I,Y$ can either be scalars, vectors or fermions.
We use the usual definition of the Mandelstam variables
\begin{align}
\sff:=s=(q_1+q_2)^2=(p-q_3)^2;	&&	u=(q_1+q_3)^2=(p-q_2)^2;\nonumber\\
	t=(q_2+q_3)^2=(p-q_1)^2=:p_I^2.
\label{eq:anomalousmandelstams}
\end{align}
where $\sff$ is the invariant mass of the two visible SM fermions and is the variable we are interested in. If one wants to derive the differential decay rate $d ~\Gamma/d~\sff$ one usually replaces one of the variables, e.g. $u$ via Mandelstam's relation
\begin{align}
u&= 2~m_f^2+m_x^2+m_y^2-t-\sff
\label{eq:mandelstam}
\end{align}
and integrates out the remaining invisible invariant mass, in this case $t$ where the upper and lower bounds are
\begin{align}
t_{\pm}=&\left( m_f^2+m_x^2+m_y^2-\sff\right)\\
& \pm\frac{\sqrt{\sff \left(\sff-4 m_f^2\right) \left(m_x^4+\left(m_y^2-\sff\right)^2-2 m_x^2
   \left(m_y^2+\sff\right)\right)}}{2 \sff}.
\end{align}
The remaining invariant mass $\sff$ has then the kinematical bounds
%\footnote{From $\sff=(q_1+q_2)^2=(m_1^2+m_2^2+2(E_1+E_2-\cos\theta |\vec{p}_1||\vec{p}_2|=(m_1^2+m_2^2)^2$, with the condition for the lower bound (fermion 1 and 2 get minimal energy) $E_1=E_2=0,\cos\theta=0$ and hence $|q_i|=m_i$, analogous for the upper bound where the fermion 1 and 2 get maximal energy and hence the other two particles minimal energy $(E_x=E_y=0)$: $\sff=(p-q_3)^2=(m_x^2+m_y^2+2(E_x E_y-\cos\theta |\vec{p}||\vec{q}_3|))^2=(m_x-m_y)^2$.}
\begin{align}
{\sff}_-= (2m_f)^2; &&
{\sff}_+ =(m_x-m_y)^2.
\label{eq:endpoints1}
\end{align}

Our interest are subsequent two body decays keeping track of the
polarisation information which is transferred via the intermediate
particle $I$. For this we use the narrow width approximation (NWA)
as worked out in ref.~\cite{Uhlemann:2008pm}. 
The endpoints of the invariant mass distribution $\sff$ after using NWA 
get changed due to the on-shell condition for the particle $I$  and reads
\cite{Nakamura:2010zzi}
\begin{align}
{\sff}_-=(2m_f)^2 && {\sff}_+=\frac{(m_x^2 - m_I^2) (m_I^2 - m_y^2)}{m_I^2}
 \, .
\end{align}

\section{Dimension 5 and dimension 6 operators}
\label{sec:anomalousoperators}

In this work we are interested in the question if \dimfive and \dimsix
operators can invalidate the spin analysis for new particles, e.g.\
by increasing the highest power of $s_{ff}$. This is only possible
if there are momentum dependent interactions and, thus, we will
restrict ourselves to this subset. Moreover, we consider final
states containing SM fermions giving further restrictions. 

The basic operator structures are given in \cite{Buchmuller:1985jz} 
which, however, involves only the SM fields as external particles.
In our examples new particles are allowed including additional
gauge interactions with covariant derivatives
 $D_\mu=\partial_\mu-iA_\mu$ and field strength
  $F_{\mu\nu}=\left(\partial_\mu A_\nu-\partial_\nu A_\mu+\ldots\right)$.
We assume that these additional gauge groups are broken at a scale
as a new scalar $\phi$ gets a vacuum expectation value $v_\phi$
and inducing a \dimfive from a \dimsix operator via
\begin{equation}
\frac{\phi}{\Lambda^2} \to
\frac{v_\phi}{\Lambda^2}=\frac{1}{\Lambda_{eff}} \,.
\end{equation}
Having this in mind we get two classes of operators: 
 fermion-fermion-vector (f-f-V) and  fermion-fermion-scalar (f-f-S).
For the (f-f-V) interactions we have
\begin{subequations}
\begin{align}
\mathcal{L}_{D4}^g=&A_\mu\overline{\psi_1}\gamma^\mu\left(g_lP_l+g_rP_r\right)\psi_2+h.c.\\
\mathcal{L}_{D5}^a=&\frac{1}{\Lambda_a}\left(\overline{\psi_1}\sigma^{\mu\nu}\left(a_lP_l+a_rP_r\right)\psi_2 F_{\mu\nu}+h.c.\right)\label{eq:sigmalagrange}\\
\mathcal{L}_{D6}=&\frac{1}{\Lambda^2}\left(\overline{(D_\mu\psi_1)}\left(b_lP_l+b_rP_r\right)\psi_2 D^\mu\phi+h.c.\right)\nonumber\\ & \rightarrow\mathcal{L}_{D5}^b=\frac{1}{\Lambda_b}\left(\overline{(\partial_\mu\psi_1)}\left(b_lP_l+b_rP_r\right)\psi_2 (-iA^\mu)+h.c.\right)&\\
\mathcal{L}_{D6}=&\frac{1}{\Lambda^2}\left(\overline{\psi_1}\left(c_lP_l+c_rP_r\right)D_\mu\psi_2 D^\mu\phi +h.c.\right)\nonumber\\ & \rightarrow\mathcal{L}_{D5}^c=\frac{1}{\Lambda_c}\left(\overline{\psi_1}\left(c_lP_l+c_rP_r\right)\partial_\mu\psi_2 (-iA^\mu) +h.c.\right)
\end{align}
\end{subequations}
and for the (f-f-S) interactions
\begin{subequations}
\begin{align}
\mathcal{L}_{D4}^n&=\left(\phi~\overline{\psi_1}\left(n_lP_l+n_rP_r\right)\psi_2+h.c.\right)	\\
\mathcal{L}_{D5}^x&=\frac{1}{\Lambda_x}\left(\overline{\psi_1}\gamma^\mu \left(x_lP_l+x_rP_r\right)\psi_2 D^\mu\phi+h.c.\right)	&	\\
\mathcal{L}_{D5}^y&=\frac{1}{\Lambda_y}\left(\overline{(D_\mu\psi_1)}\gamma^\mu \left(y_lP_l+y_rP_r\right)\psi_2\phi+h.c.\right)&	
\label{eq:dim5scalar}\\
\mathcal{L}_{D5}^z&=\frac{1}{\Lambda_z}\left(\overline{\psi_1}\gamma^\mu \left(z_lP_l+z_rP_r\right)(D_\mu\psi_2)\phi		+h.c.\right)
\end{align}
\label{eq:ffs}
\end{subequations}
Obviously not all of them are independent, e.g.\ by partial
integration one can transfer one of the derivatives to the other
two fields. In principle one could also consider \dimsix operators
for the (f-f-S) case but as we will discuss below, no additional features
will occur in such a case.

\section{Results}\label{sec:resultsanomalous}

We will first the discuss the impact of operators containing 
two fermions and scalar as here the analytical formulas are
rather compact. Then we will turn to decays involving also
vector bosons.

\subsection{A simple example}

We start with a simple example, namely the case where $X$ and $Y$
are fermions and $I$ a scalar and we denote this
case for later use by $(FF)_S$. In this case that there are no
spin correlations between the SM fermions, the matrix element,
including the anomalous couplings due to the  \dimfive operators,
is of the form
\begin{align}  
\mathcal{M}_{D4,D5}&\propto \left(\overline{u}(f)(1+\lambda_1 k_1\!\!\!\!/)u(x)\right)\left(\overline{u}(y)(1+\lambda_2 k_2\!\!\!\!/)v(\bar{f})\right)
\end{align} 
where we have used a combinations of $\mathcal{L}_{D5}^y$ and
$\mathcal{L}_{D5}^z$, e.g.\ the last two operators in
eq.~(\ref{eq:ffs}). The $k_i$ are weighted sum of the
fermion momenta, e.g.\ $k_1=\alpha_1 p + \beta_1 q_1$ and
$k_2=\alpha_2 q_2 + \beta_2 q_3$, and $\lambda_i$ are
measures of the relative importance of the \dimfive with respect
to the \dimfour operators. 
 Here we have put for simplicity scalar couplings
only. As all fermions are on-shell, this immediately implies that
one can use the equation of motion and replace in the $k_i\!\!\!\!/$ 
the momenta by
the corresponding fermion masses. This immediately implies that no
additional power of $s_{ff}$ occurs and we get
\begin{equation}
\frac{d~\Gamma}{d \sff} = A\cdot\sff^0 = A.
\end{equation}
In the case that $\mathcal{L}_{D5}^x$ is involved one has
to use the momentum conservation of the vertices to express
the momentum of the scalar by combinations of the fermion momenta.
One can also show easily in this case that using of the equations
of motion gives only additional mass terms.

The more complicated case is that $X$ and $Y$ are scalars and $I$
a fermion as in the case $I$ relates the polarisation of the two
SM fermions and we denote this kind of decay for later use as
$(SS)$. However, in this case also an explicit calculation
shows that no additional power of   $s_{ff}$ occurs and we get
\begin{equation}
\frac{d~\Gamma}{d \sff} = A + B \cdot\sff
\end{equation}
The result for the coefficients $A$ and $B$ is given in the
appendix. One can show along the same lines, that operators of the
form
\begin{align}
\mathcal{L}_{D6}&=\frac{1}{\Lambda^2}\left(\overline{\psi_1} 
\psi_2 D^\mu D_\mu\phi+h.c.\right)	
 + \frac{1}{\Lambda^2}\left(\overline{\psi_1}  (D^\mu D_\mu \psi_2)
 \phi+h.c.\right)&	
\end{align}
also do not give any contributions to these two decay chains. 
As we only wanted to indicate
the principle structure of the couplings
we did not write any chiral couplings as they are not important 
in this context.

\subsection{Decays involving vector bosons}

\begin{figure}[t]\centering
\includegraphics[width=1.5cm]{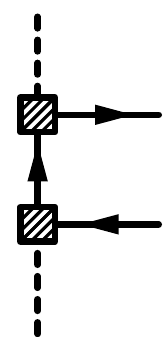}
\hspace{3ex}
\includegraphics[width=1.5cm]{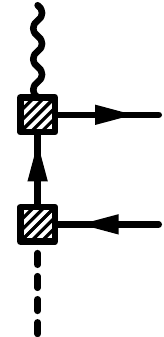}
\hspace{3ex}
\includegraphics[width=1.5cm]{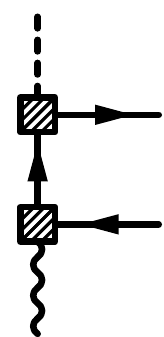}
\hspace{3ex}
\includegraphics[width=1.5cm]{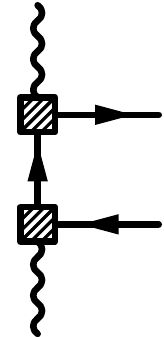}
\hspace{3ex}
\includegraphics[width=1.5cm]{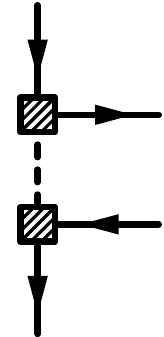}
\hspace{3ex}
\includegraphics[width=1.5cm]{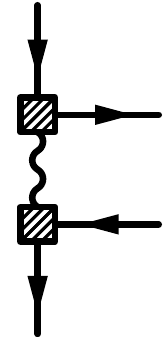}
\caption{Feynman diagrams for the decays involving intermediate fermions and a scalar/vector boson. The boxes denote combinations
of \dimfour and \dimfive couplings.}
\label{fig:fermiintermed}
\end{figure}

Here we have four cases, three where the intermediate particle
is a fermion
\begin{subequations}
\begin{align}
S\rightarrow f\bar{f}V	&&(SV)\\
V\rightarrow f\bar{f}S	&&(VS)\\
V\rightarrow f\bar{f}V	&&(VV)
\end{align}
\end{subequations}
yielding a  differential decay rate of the form
\begin{align}
\frac{d~\Gamma}{d\sff}&= A+B\cdot s_{ff}^1
\label{eq:fermiintermed}
\end{align} 
where $A,B$ contain the dependence on masses and couplings  involved. 
The Feynman graphs for the corresponding decays are the second to
fourth  in Fig.~\ref{fig:fermiintermed}.
The final decay process shown in Fig. \ref{fig:fermiintermed}  contains an
intermediate vector boson
\begin{align}
F\stackrel{V}{\longrightarrow} f\bar{f}F	&&(FF)_V
\end{align}
yielding a partial width of the form
\begin{align}
\frac{d~\Gamma}{d \sff}&= A +B\cdot \sff^1+C\cdot \sff^2
\label{eq:anomalousbosonexpectation}
\end{align}
$A,B,C$ are again functions of the involved couplings and the masses. 
In all cases we have already anticipated that no additional
powers of $\sff$ are induced. The only effect of the anomalous
couplings is to change somewhat the relative size of the coefficients
$A$, $B$ and $C$.

We sketch the main results for the case of $(VV)_F$ as this is the
one where one might expect the highest power of momenta and thus
the highest power of $\sff$.
The amplitude for this decay  leads to the following matrix element
\begin{flalign}
&\mathcal{M}_{(VV)}=\epsilon^\mu(p)\epsilon^{\nu*}(q_3)	\overline{u}(q_1)\left(\Gamma_{1~V}^{D4}+\Gamma_{1~V}^{D5}\right)_{\mu}\left(\frac{p\!\!\!\!/-q_1\!\!\!\!/+m_I}{(p_I^2-m_I^2)}\right)\left(\Gamma_{2~V}^{D4*}+\Gamma_{2~V}^{D5*}\right)_{\nu}v(q_2)&&\label{eq:decayvv}
\end{flalign}
where we have indicated the parts stemming from \dimfour (\dimfive) 
operators by $\Gamma_{i~V}^{D4}$ ($\Gamma_{i~V}^{D5}$), where $i=1,2$ indicate the first or second vertex in the decay.
For the squared matrix element we get the following structure
\begin{align}
\sum_{pols}|\mathcal{M}_{(VV)_F,D4}|^2 &\sim \sum_{pols}\left(\overline{u}(q_1)\gamma_\mu p_I\!\!\!\!/\gamma_{\nu} v(q_2)\right)\left(\overline{v}(q_2)\gamma_{\nu'}p_I\!\!\!\!/\gamma_{\mu'} u(q_1)\right)\nonumber\\
&\quad\quad\times	\epsilon^{\mu}(p)\epsilon^{\nu*}(q_3)\cdot\epsilon^{\mu'*}(p)\epsilon^{\nu'}(q_3)\nonumber
\\
&\sim Tr\left[q_1\!\!\!\!/\gamma_\mu p_I\!\!\!\!/\gamma_\nu q_2\!\!\!\!/\gamma_{\nu'} p_I\!\!\!\!/\gamma_{\mu'}\right] \cdot \frac{p^\mu p^{\mu'}}{m_x^2}\cdot \frac{q_3^\nu q_3^{\nu'}}{m_y^2}\nonumber\\ &
\sim -4\left(m_x^2-2~p\cdot q_1\right)^2(m_x^2 ~q_1\cdot q_2-2 ~p\cdot q_1~p\cdot q_2)
\end{align}
using only \dimfour contributions and where we have omitted all chiral 
couplings as they are not important at this stage. 
After taking the NWA, the Mandelstam variables become
\begin{subequations}
\begin{align}
t&=(p-q_1)^2=(q_2+q_3)^2=m_I^2\label{eq:nwamandelstamss}
 \\ u&=(q_1+q_3)^2=(p - q_2)^2=2m_f^2+m_x^2+m_y^2-\sff-m_I^2
\label{eq:nwamandelstamst}
\end{align}
\label{eq:nwamandelstams}
\end{subequations}
Therefore, in NWA we can only get a $\sff$ term from the scalar products
\begin{align}
p\cdot q_2\rightarrow \sff+const.;	&&	q_1\cdot q_3\rightarrow \sff+const.
\label{eq:anomsffterms}
\end{align}
while the remaining scalar products just give a mass squared and hence 
constant contributions. Therefore we get as claimed a linear
function in $\sff$ after taking the NWA. 
Similarly one finds
the $\sff$ dependence in the cases $(SV)$ and $(VS)$ including only
the \dimfour operators. It turns out that the 
impact of the
\dimfive operators on these decays is the same as for the $(FF)_V$
case and, thus, we will present first the main facts for this decay
using \dimfour operators only and discuss the \dimfive operators
for all four decays taking the $(VV)$ case as an example.

In the case of $(FF)_V$, the most involved one, we get 
\begin{subequations}
\begin{align}
\mathcal{M}_{(FF)_V}&=	\left(\overline{u}(q_1)\left(\Gamma_{1V}^{D4}+\Gamma_{1V}^{D5}\right)_\mu u(p)\right)\cdot\left(\overline{u}(q_3)\left(\Gamma_{2V}^{D4}+\Gamma_{2V}^{D5}\right)_\nu v(q_2)\right)\nonumber
\\
&\quad\quad
\times\left(\frac{g^{\mu\nu}-p_I^\mu p_I^\nu/m_I^2}{p_I^2-m_I^2}\right)\label{eq:decayffv}
\end{align}
\end{subequations}
indicating again the parts coming from \dimfour (\dimfive)
operators by $\Gamma_{1V}^{D4}$ ($\Gamma_{1V}^{D5}$). For
the squared amplitude we obtain for the part stemming from
the  \dimfour operators
\begin{align}
&\sum_{pols}|\mathcal{M}_{(FF)_V,~D4}|^2\sim\nonumber\\ &\sim\sum_{pols}\left(\overline{u}(q_1)\gamma_\mu u(p)\overline{u}(p)\gamma_\nu u(p_1)\right)\left(\overline{u}(q_3)\gamma_{\nu'}v(q_2)\overline{v}(q_2)\gamma_{\mu'}u(q_3)\right)
\nonumber \\
& \quad\quad\hspace*{2mm} \times  (g^{\mu\nu}-\frac{p_I^\mu p_I^\nu}{m_I^2})(g^{\mu'\nu'}-\frac{p_I^{\mu'}p_I^{\nu'}}{m_I^2})\nonumber\\
&\sim Tr\left[q_1\!\!\!\!/\gamma_\mu p\!\!\!\!/\gamma_\nu\right]Tr\left[q_2\!\!\!\!/\gamma_{\nu'}q_3\!\!\!\!/\gamma_{\mu'}\right]\cdot  (g^{\mu\nu}-\frac{p_I^\mu p_I^\nu}{m_I^2})(g^{\mu'\nu'}-\frac{p_I^{\mu'}p_I^{\nu'}}{m_I^2})\nonumber\\ &
\sim 16(-g^{\mu\mu'} p\cdot q_1+p^\mu q_1^{\mu'}+p^\mu q_1^{\mu'})(-g^{\nu\nu'} q_2\cdot q_3+q_2^\nu q_3^{\nu'}+q_2^\nu q_3^{\nu'})
\nonumber \\ & \quad\quad\hspace*{2mm}
\times  (g^{\mu\nu}-\frac{p_I^\mu p_I^\nu}{m_I^2})(g^{\mu'\nu'}-\frac{p_I^{\mu'}p_I^{\nu'}}{m_I^2})
\label{eq:vvexample0}
\end{align}
In comparison to the case $(VV)$ we have the additional contributions 
of $p^\mu q_1^{\mu'}\ldots$ which can then be contracted with the momenta
combination of the trace of the second fermion line. These contractions 
give new momenta combinations and one can see that terms of the form 
$(q_1\cdot q_2)~ (p\cdot q_3)$ arise which give according to 
eq.~(\ref{eq:anomsffterms})
 a $\mathcal{O}(\sff^2)$ contribution which is the 
characteristic for intermediate vector bosons.
Note, that $(p\cdot q_3) = \sff +const.$ according to
 eq.~(\ref{eq:anomalousmandelstams}).

In both cases the couplings induced by the \dimfive operator
have in the most general case the form
\begin{align}
\Gamma_{iV}^{D5\mu}&= A_i\cdot \sigma^{\mu\alpha}k_{V~\alpha}
+B_i ~ k_{1i}^\mu+C_i ~k_{2i}^\mu \,\, (i=1,2)
\end{align}
where $k_V$ is the momentum of the vector boson, and $k_{1i}$
($k_{2i}$) are linear combinations of the particle momenta at
the first (second) vertex.
Plugging this in eq.~(\ref{eq:decayvv}) and considering the part of the
matrix element squared with the anomalous coupling squared, as this
gives potentially the highest power in $\sff$, we find
\begin{flalign}
&\sum_{pols}|\mathcal{M}_{VV,~D5}|^2\sim\nonumber&&\\ &\sim \sum_{pols}\left(\overline{u}(q_1) (A_1\sigma^{\mu\alpha}p_{\alpha}+B_1 p_I^\mu +C_1 q_1^\mu) p_I\!\!\!\!/ (A_2\sigma^{\nu\alpha}q_{3~\alpha}+B_2 p_I^\nu +C_2 q_2^\nu) v(q_2)\right)\nonumber &&\\
& \quad\quad\times 
\left(\overline{u}(q_1) (A_1\sigma^{\mu'\alpha}p_{\alpha}+B_1 p_I^{\mu'} +C_1 q_1^{\mu'}) p_I\!\!\!\!/ (A_2\sigma^{\mu'\alpha}q_{3~\alpha}+B_2 p_I^{\mu'} +C_2 q_2^{\mu'}) v(q_2)\right)^\dagger\nonumber &&\\ &\quad\quad
\times \underbrace{\epsilon_\mu(p)\epsilon_{\mu'}(p)^*\epsilon_\nu(q_3)\epsilon_{\nu'}(q_3)^*}_{\frac{(p_\mu p_{\mu'})}{m_x^2}\frac{(q_{3,\nu} q_{3,{\nu'}})}{m_y^2}
}.&&
\end{flalign}
We see immediately that the $\sigma^{\mu\nu}$  terms drop out as they 
are contracted
with symmetric products of momenta. The remaining products of momenta
are
\begin{equation}
p_I^{\mu} p_\mu = (p-q_1)^{\mu} p_\mu
\,\,,\,\, q_1^{\mu} p_\mu \,\,,\,\,
p_I^{\mu} q_{3,\mu} = (q_2+q_3)^{\mu} q_{3,\mu} \,\,,\,\, q_2^{\mu} q_{3,\mu}
\end{equation}
Using eq.~(\ref{eq:nwamandelstams}) we see that in NWA all of
them give only sums of masses squared but no additional powers of
$\sff$. The same reasoning can also be used to show that also in
the case of $(SV)$, $(VS)$ and $(FF)_V$ no additional powers
of $\sff$ are induced.

The main reasons, why higher dimensional operators do
not change the overall lepton and quark  spectra
of the decays, can be summarized as follows:
\begin{enumerate}
\item Additional $\sff$ dependence, which is equivalent to
 additional $\cos \theta$
dependence, can only arise through the
following products: $(q_1\cdot q_3)$  or $(p\cdot q_2)$. 
 All other give
in NWA sums of masses squared, see eqs.~(\ref{eq:nwamandelstamss})
and (\ref{eq:nwamandelstamst}).
\item The antisymmetric part, e.g.\ the $\sigma^{\mu\nu}$ part,
of the (f-f-V) coupling gets always contracted by the same momentum due 
to the polarization sum/propagator of the vector boson  
and hence gives zero.
\item The momentum dependent parts in the (f-f-V) coupling $k^\mu$ 
relate only momenta within a given vertex. In NWA the momentum
conservation at a given vertex implies that all scalar products
of momenta can be expressed either as masses squared or as $t=m_I^2$.
\item Momenta contracted with gamma-matrices yield only masses 
after using the  Dirac equation.
\end{enumerate}
Therefore, also \dimfive operators where fermions are
coupled to vector bosons do not change the highest power in $\sff$.

We have checked that the same reasoning also applies for
\dimsix operators.
Higher than \dimsix operators should play no role as latest
at this stage higher order corrections due to emission of
gluons and photons become more important.

\subsection{Numerical example}

\begin{figure}\centering
\includegraphics[width=0.49\textwidth]{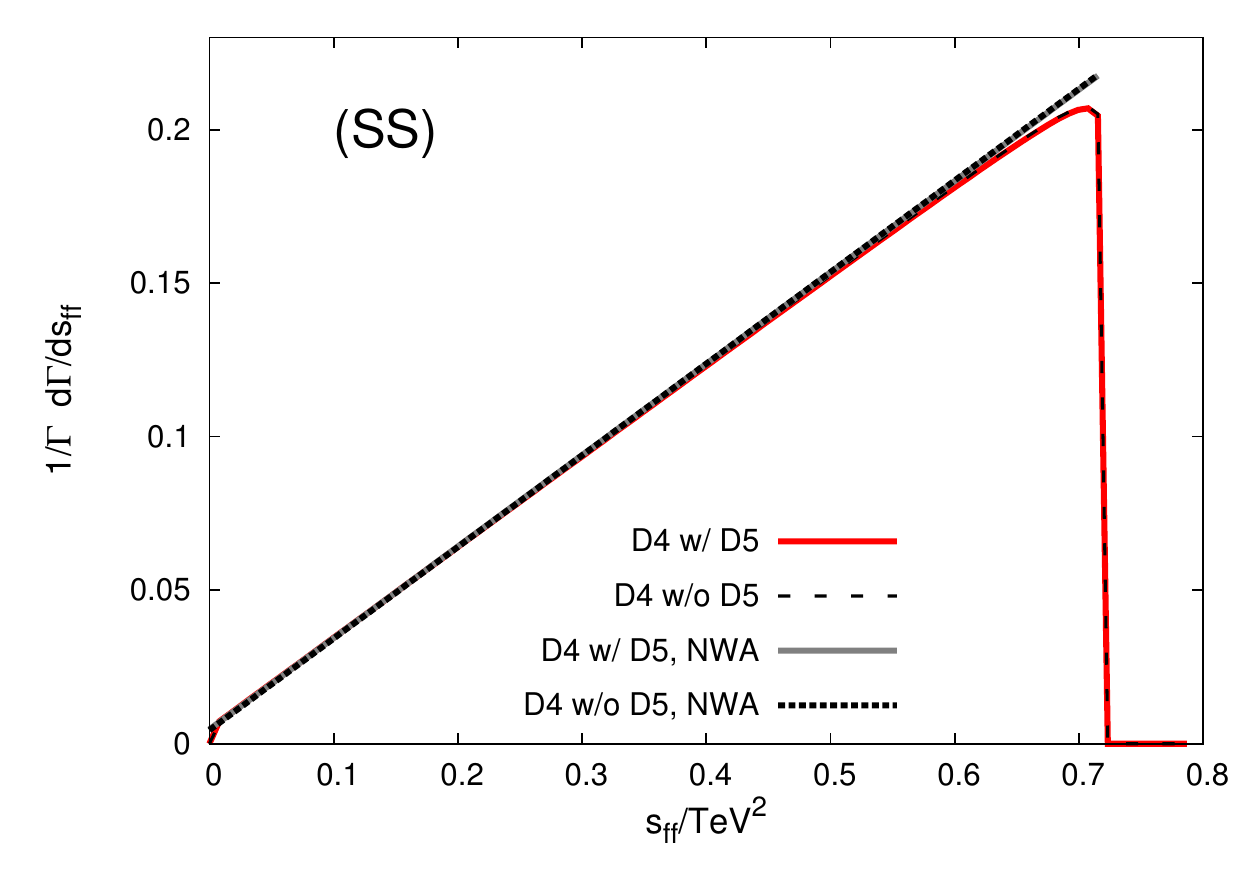}
\includegraphics[width=0.49\textwidth]{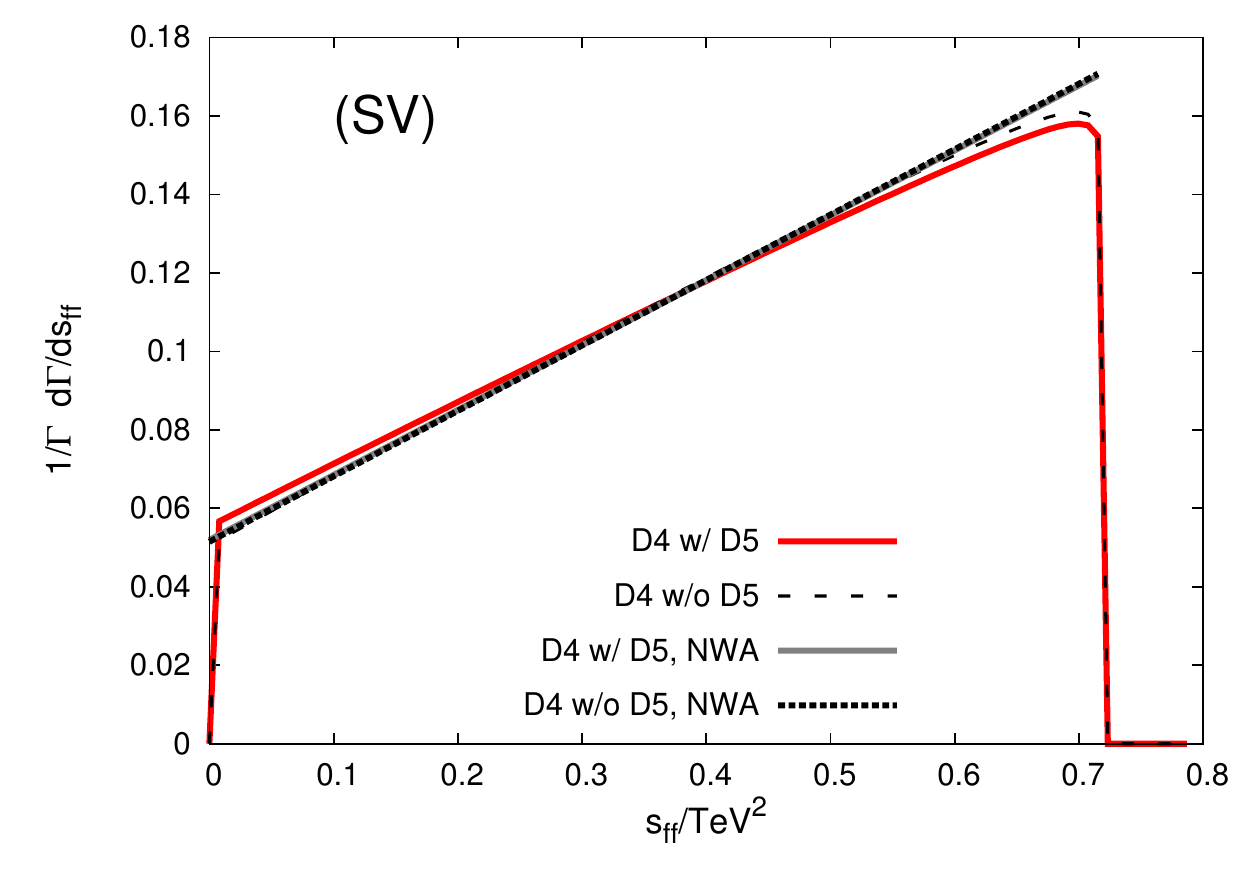}
\includegraphics[width=0.49\textwidth]{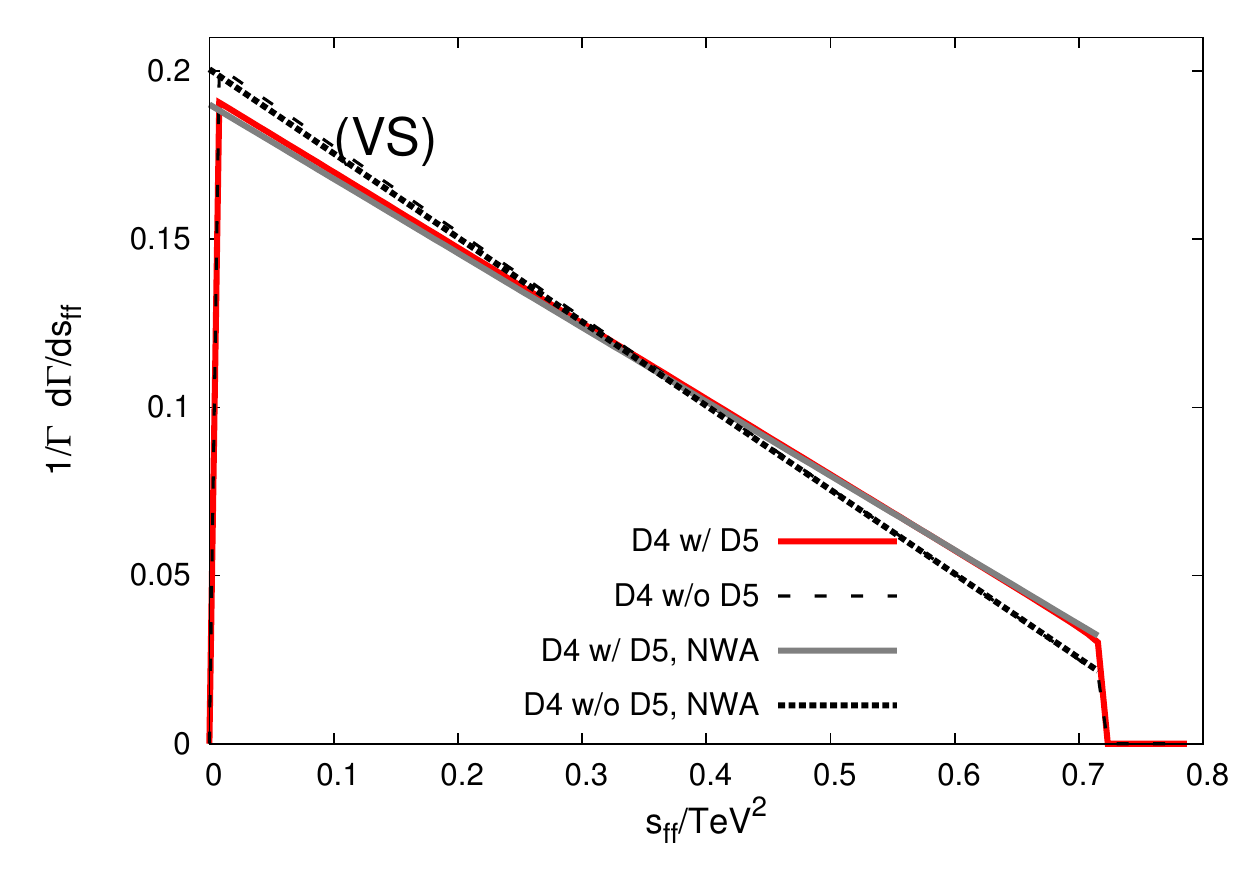}
\includegraphics[width=0.49\textwidth]{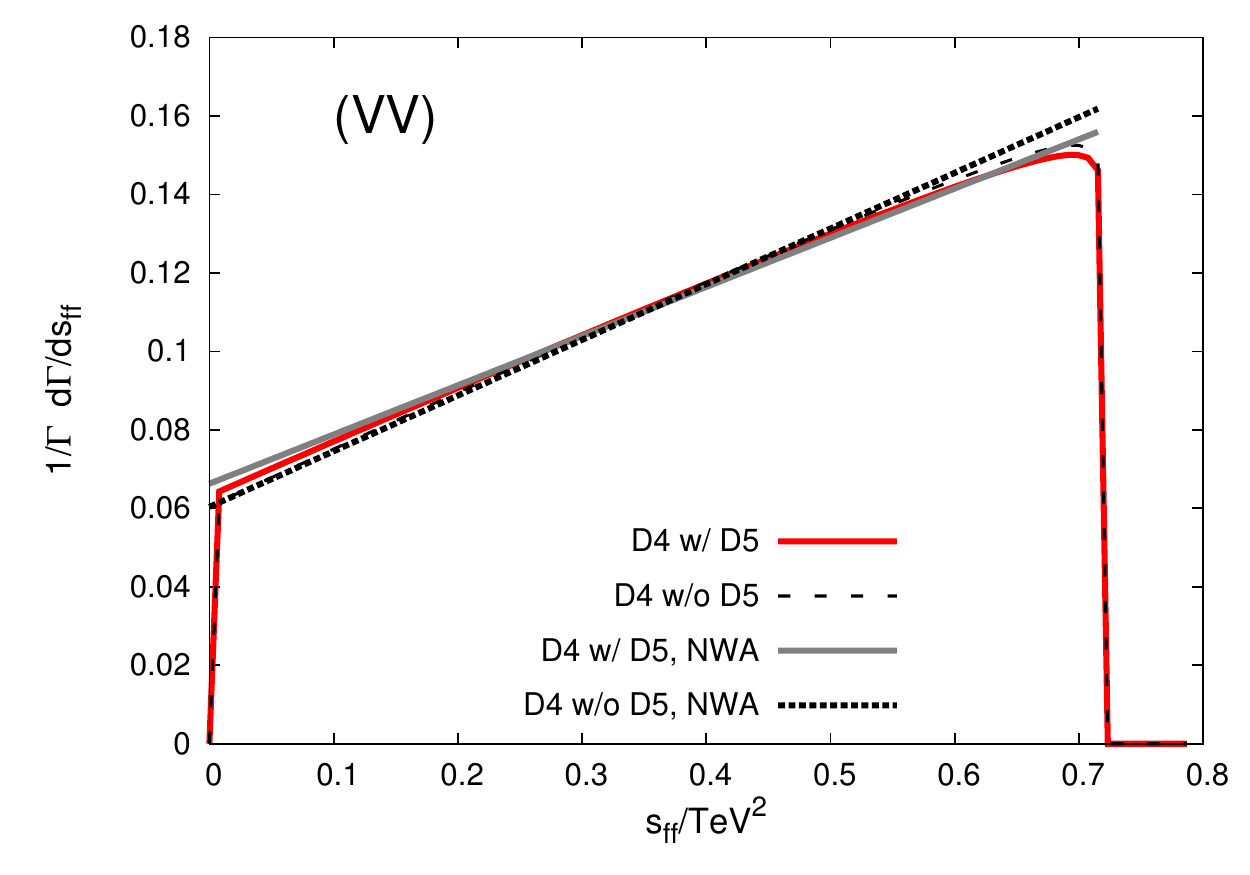}
\includegraphics[width=0.49\textwidth]{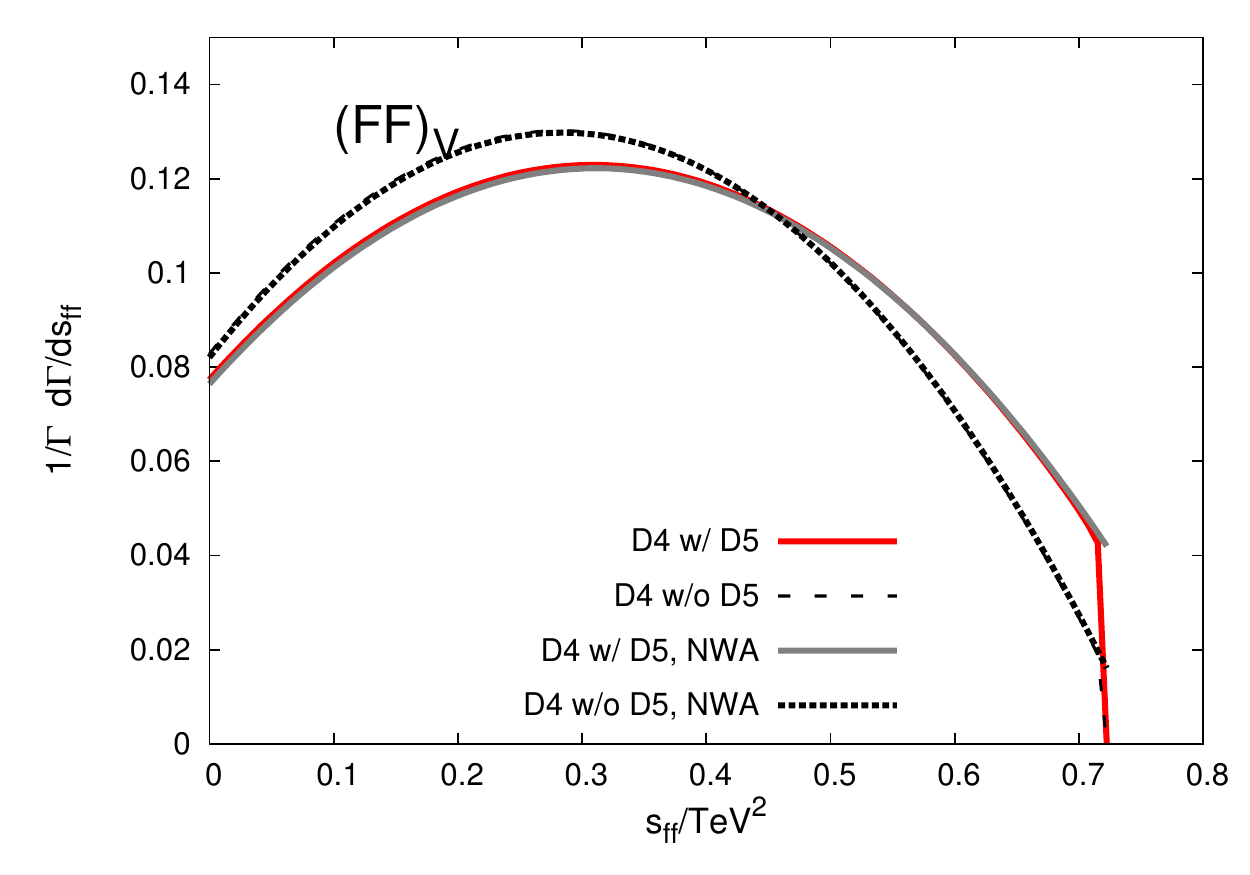}
\includegraphics[width=0.49\textwidth]{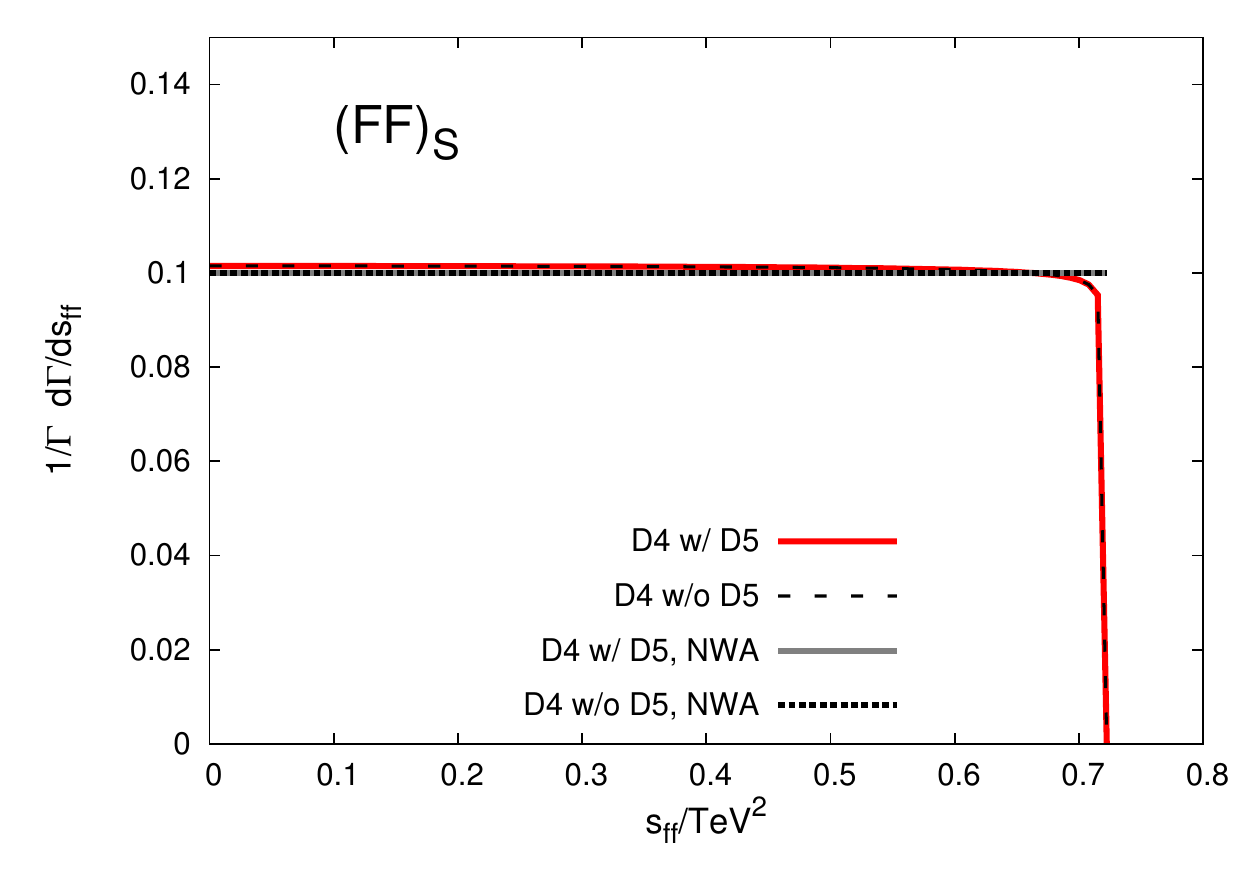}
\caption{Differential decay rate over the invariant mass $\sff$ for  
\dimfour couplings only (black) and including additional \dimfive 
(red/grey) couplings for intermediate fermions/bosons (first two row/last 
row). The lines correspond to:
full red lines: exact spectra including \dimfour and \dimfive operators,
dashed lines: exact spectra taking \dimfour operators only,
full grey lines: spectra including \dimfour and \dimfive operators using NWA,
dotted lines: exact spectra taking \dimfour operators only using NWA.}
\label{fig:anomalousplot}
\end{figure}

We have checked numerically several examples to test the
quality of our reasoning in the previous sections. 
As a random example,
we show for all six decays the differential decay rate 
$\frac{1}{\Gamma} \frac{d\Gamma}{d \sff}$ for the 
\dimfour and \dimfive result in fig.~\ref{fig:anomalousplot},
The masses and the couplings are chosen as  
\begin{align}
m_x=1~TeV && m_y=0.15~TeV	&&	m_f=m_{\bar{f}}=0 \nonumber \\
m_I=0.4~TeV&& \Lambda=3~TeV \nonumber \\
 n_l =1, n_r = 0.1; && g_l = 1, g_r = 0.1;&& \\
 a_l = 1,  a_r = 0.3; && b_l = 0.4, b_r = 0.5; && c_l = 0.8, c_r = 0.7;
\nonumber \\
 x_l = 0.4,    x_r = 0.3; && y_l = 0.5, y_r = 0.3; && z_l = 1, z_r = 0.1
 \,.\nonumber
\label{eq:anomalousparameters}
\end{align}

We show for each decay the complete \dimfour (black dashed) and \dimfive
 distribution (red solid)  and for comparison the NWA results (\dimfour, NWA: black dotted; \dimfive, NWA: grey solid). For all decays, the NWA result is a very good approximation for the complete decays including 
 off-shell effects. The main differences are at the kinematical
endpoints. The influence of the \dimfive operators in this mass/coupling scenario is the largest in the $(FF)_V$ followed by the $(VS)$ decay. The
first one is because in this case we have the highest
power in $\sff$, see eq.~(\ref{eq:maximalorder}), and is a quite 
generic feature. The second case depends much stronger on the couplings
and masses of the scenarios considered.  For completeness
we note, that for $(FF)_S$  no changes are visible when comparing
the cases with and without \dimfive operators, because we
show the normalized distributions.

%%%%%%%%%%%%%%%%%%%%%%%%%%%%%%%%%%%%%%%%%%%%%%%%%%%%%%%%%%%%%%%%%%%%%%%%%%%%%
%%%%%%%%%
%%%%%%%%%%%%%%%%%%%%%%%%%%%%%%%%%%%%%%%%%%%%%%%%%%%%%%%%%%%%%%%%%%%%%%%%%%%%%

\section{Discussion and conclusions}
\label{sec:discussion}
%%%%%%%%%%%%%%%%%%%%%%%%%%%%%%%%%%%%%%%%%%%%%%%%%%%%%%%%%%%%%%%%%%%%%%%%%%%%%
%%%%%%%%%
%%%%%%%%%%%%%%%%%%%%%%%%%%%%%%%%%%%%%%%%%%%%%%%%%%%%%%%%%%%%%%%%%%%%%%%%%%%%%
The determination of spins of new particles is an important
task at the LHC once an extension of the SM is discovered. In this
way one can also discriminate between model classes, e.g.\ between
supersymmetry and extra dimensions. In case that there are two subsequent
two-body decays involving two SM fermions one can show that the
spin of the intermediate particle reflects itself as in the
highest power of $\sff$ in the differential decay width 
$d \Gamma / d \sff$.

There are two main theoretical uncertainties in this kind of
considerations: real emission of photons and gluons in the decays.
These can be calculated once the quantum numbers of the new particles
are know. The second class are higher than \dimfour operators
which are induced by particles close to the LHC reach. In this paper
we have shown that such operators do not induce higher powers of
$\sff$ in the partial width and thus the existing analyses are valid
even if such operators are numerically important. However, their
presence can change the exact form of the slope considerably as we have
seen in a concrete example. This potentially affects the determination
of the underlying parameters.

\normalfont
\section*{Acknowledgements}
W.P. thanks IFIC/C.S.I.C. for hospitality during an extended stay.
This work has been supported by the German Ministry of Education and
Research (BMBF) under contract no.\ 05H09WWE. W.P.\ is partially  
supported by the Alexander von Humboldt Foundation.

\section*{Appendix}
As an example for the NWA result of the decays including anomalous couplings, we give here the decay width for $S\rightarrow f\bar{f}S$. We used the following short forms for the couplings from eq.~\ref{eq:ffs} and the masses used 
\begin{align*}
\tau_I=\frac{m_I}{m_X};	&& \tau_y=\frac{m_y}{m_x};	&& 1/\Lambda_i=R_i=\frac{R'_i}{m_x};&&	\\
g_V=g_r+g_l;&& g_A= g_r-g_l; 	&&
n_V=n_r+n_l;&&n_A= n_r-n_l;
\end{align*}
and analogous for $a_{l/r},b_{l/r},c_{l/r},x_{l/r},y_{l/r},z_{l/r}$. The NWA result in terms of these is: 
\begin{align*}
\frac{d~\Gamma}{d~\sff}=\frac{1}{(2\pi)^3}\frac{1}{32m_X}\frac{1}{m_I\Gamma_I}\left((A_{D4}+A_{D5})+\sff\cdot(B_{D4}+B_{D5})\right)
\end{align*}
with
\begin{flalign*}
&A_{D4}=&&2 m_X^2 \left(1-\tau_I^2\right) (\tau_I^2-\tau_Y^2) 
\left(\left(g_A^2 + g_V^2\right)\left(n_A^2+n_V^2\right) +4 g_A g_V n_A n_V 
\right);
  \nonumber&&\\ 
&B_{D4}=&& -16 g_A ~ g_V~ n_A ~n_V ~ \tau_I^2;
&&\\
&A_{D5}=&&
-2 m_X^2 \tau_I^2 \left(\tau_I^2-1\right) (\tau_I^2-\tau_Y^2)\times\left[a_A^2 R^{'2}_a \Bigl(n_A^2+n_V^2
\right.&&\\&&& \left.\left.
+\tau_I^2
   \left(R^{'2}_x \left(x_A^2+x_V^2\right)-2 R'_x R'_z (x_A z_A+x_V z_V)
+R^{'2}_z \left(z_A^2+z_V^2\right)\right)\right)\right.
&&\\ &&&\left.
+2 a_A R'_a
   \left(\tau_I^2 \Bigl(2 (a_V R'_a-c_V R'_c) (R'_x x_A-R'_z z_A) (R'_x x_V-R'_z z_V)
\right.\right.&&\\&&&\left.\left.\left.
-c_A R'_c \left(R^{'2}_x
   \left(x_A^2+x_V^2\right)-2 R'_x R'_z (x_A z_A+x_V z_V)+R^{'2}_z \left(z_A^2+z_V^2\right)\right)\right)
\right.\right.&&\\&&&\left.
+2 a_V n_A n_V
   R'_a-R'_c \left(c_A \left(n_A^2+n_V^2\right)+2 c_V n_A n_V\right)\Bigr)
\right.&&\\&&&\left.
+\tau_I^2 \left[a_V^2 R^{'2}_a \left(R^{'2}_x
   \left(x_A^2+x_V^2\right)-2 R'_x R'_z (x_A z_A+x_V z_V)+R^{'2}_z \left(z_A^2+z_V^2\right)\right)
\right.\right.&&\\&&&
-2 a_V R'_a R'_c \Bigr(2 c_A
   (R'_x x_A-R'_z z_A) (R'_x x_V-R'_z z_V)
&&\\&&&\left.\left.\left.
+c_V \left(R^{'2}_x \left(x_A^2+x_V^2\right)-2 R'_x R'_z (x_A z_A+x_V
   z_V)+R^{'2}_z \left(z_A^2+z_V^2\right)\right)\right)
\right.\right.&&\\&&&\left.\left.
+R^{'2}_c \left((c_A^2+c_V)^2 \left(R^{'2}_x \left(x_A^2+x_V^2\right)-2 R'_x R'_z (x_A z_A+x_V
   z_V)+R^{'2}_z \left(z_A^2+z_V^2\right)\right)
\right.\right.\right.&&\\&&&
+4 c_A c_V (R'_x x_A-R'_z z_A) (R'_x x_V-R'_z z_V)\Bigr)\Bigr]
&&\\&&& 
+a_V^2 n_A^2
   R^{'2}_a+a_V^2 n_V^2 R^{'2}_a-4 a_V c_A n_A n_V R'_a R'_c
&&\\&&&
-2 a_V c_V n_A^2 R'_a R'_c-2 a_V c_V n_V^2 R'_a
   R'_c
&&\\&&& 
+c_A^2 n_A^2 R^{'2}_c+c_A^2 n_V^2 R^{'2}_c+4 c_A c_V n_A n_V R^{'2}_c+c_V^2 n_A^2 R^{'2}_c+c_V^2 n_V^2
   R^{'2}_c
&&\\&&& \left.
+(g_A^2 +g_v^2)\left(R^{'2}_x \left(x_A^2+x_V^2\right)-2 R'_x R'_z (x_A z_A+x_V z_V)+R^{'2}_z \left(z_A^2+z_V^2\right)\right)
\right.
&&\\&&&
+4 g_A
   g_V (R'_x x_A-R'_z z_A) (R'_x x_V-R'_z z_V)\Bigr];
 &&\\ &
B_{D5}=&& -16  \tau_I^4 \left((a_A R'_a-c_A R'_c) (a_V R'_a-c_V R'_c) \left(n_A n_V
\right.\right.&& \\ &&&\left.\left.
+\tau_I^2 (R'_x x_A-R'_z z_A)
   (R'_x x_V-R'_z z_V)\right)
\right.&&\\&&&\left.
+g_A g_V (R'_x x_A-R'_z z_A) (R'_x x_V-R'_z z_V)\right).&&
\end{flalign*}

\end{document}